\documentclass[twocolumn,showpacs,preprintnumbers]{revtex4}%
\usepackage{amsfonts}
\usepackage{amssymb}
\usepackage{amsmath}
\usepackage{graphicx}
\usepackage{dcolumn}
\usepackage{bm}%
\providecommand{\U}[1]{\protect\rule{.1in}{.1in}}
\begin{document}
\title{The phase sensitivity of an SU(1,1) interferometer with homodyne detection }
\author{Dong Li$^{1}$, Chun-Hua Yuan$^{1,*}$, Z. Y. Ou$^{1,2}$, and Weiping
Zhang$^{1,\dag}$}
\affiliation{$^{1}$Quantum Institute for Light and Atoms, Department of Physics, East China
Normal University, Shanghai 200062, P. R. China}
\affiliation{$^{2}$Department of Physics, Indiana University-Purdue University
Indianapolis, 402 North Blackford Street, Indianapolis, Indiana 46202, USA}
\date{\today }

\begin{abstract}
We theoretically study the phase sensitivity of the SU(1,1) interferometer
with a coherent state in one input port and a squeezed-vacuum state in the
other input port using the method of homodyne. We find that the the relative
phase sensitivity is not always better with increasing the input squeezed
strength or increasing the input coherent light intensity. We give out the
optimal condition to make the relative phase sensitivity to reach the
Heisenberg limit, where the parameter requirements can be realized with
current experiment technology.

\end{abstract}

\pacs{42.50.St, 07.60.Ly, 42.50.Lc, 42.65.Yj}
\maketitle

The interferometer is a fundamental physical apparatus that has been
implemented using photons, neutrons \cite{Sakurai}, electrons \cite{Ji}, and
atoms \cite{Berman}. It has broad applications that a number of different
interferometer configurations have been proposed to measure the orbital
angular momentum of the photon \cite{Jonathan}, to study the berry phase
\cite{Erik}, to encode qubits \cite{Pan}. The optical interferometer is a very
useful and flexible measuring tool for phase estimation
\cite{Caves81,Xiao,Giovannetti,NOON,Dowling08,Ou12,Ou13}. The precision with
which phase shifts can be determined in an interferometer is limited by shot
noise or the standard quantum limit (SQL), $\triangle\phi_{\text{SQL}}%
\sim1/\sqrt{N}$. Recent designed quantum procedures make it possible to reduce
the classic noise to the level where the quantum noise becomes dominant that
beat the SQL and reach the Heisenberg Limit (HL) $\triangle\phi_{\text{HL}%
}\sim1/N$ \cite{Caves81,Xiao,Giovannetti,NOON,Dowling08,Ou12,Ou13}.

To beat the SQL, part of the research work based on the SU(2) type
Mach-Zehnder interferometer (MZI) consider how to improve the input state.
Different nonclassical input states have been suggested, such as, coherent
$\otimes$ squeezed-vacuum light as input of a MZI suggested by Caves
\cite{Caves81} and realized by Xiao \emph{et. al.} \cite{Xiao}, NOON states
\cite{NOON}, etc. Another line of research follows the trail of improving
measurement methods. Some measurement schemes have been proposed to improve
the phase sensitivity, such as, parity measurements \cite{Campos} which can be
used to beat the HL \cite{Anisimov}, photon counts of two output ports
\cite{Luza}, etc. Another part of research work pay attention to how to change
the structure or hardware of the interferometer to enhance the phase
sensitivity \cite{Yurke,Jacobson,Ou97,Ou12,Ou13}. For example, the nonlinear
elements were introduced in the linear interferometers. In 1986, such a class
of interferometers introduced by Yurke \emph{et. al.} \cite{Yurke} is
described by the group SU(1,1)-as opposed to SU(2). This class of
interferometers may be realized by replacing the 50-50 beam splitters with
four-wave mixers (FWMs) or parameter amplifiers in a traditional MZI
\cite{Yurke,Ou12,Ou13,Plick}. Recently, our group firstly realized it in
experiment with FWMs acting as beam splitters to split and recombine the
incoming optical fields \cite{Jing}. The phase sensitivity of the SU(1,1)
interferometer with coherent state input has been studied using the intensity
detection in detail \cite{Plick,Marino}. The coherent light together with a
squeezed vacuum injecting the SU(1,1) interferometer to break through the SQL
has been proposed \cite{Ou12}. Here, we further study the phase sensitivity by
using coherent $\otimes$ squeezed-vacuum light as input of a SU(1,1)
interferometer in detail.

\begin{figure}[ptb]
\centerline{\includegraphics[scale=0.65,angle=0]{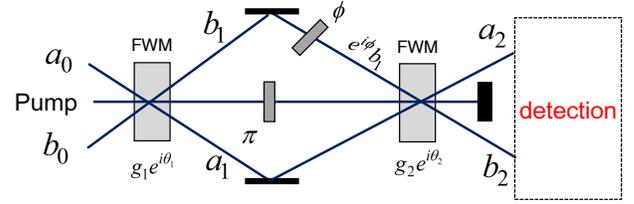}} \caption{(Color
online) The schematic diagram of SU(1,1) interferometer. Two FWMs take the
place of two beam splitters in the traditional Mach-Zehnder interferometer.
$g_{1}$ ($g_{2}$) and $\theta_{1}$ ($\theta_{2}$) describe the strength and
phase shift in the FWM process $1$ ($2$), respectively. $a_{i}$ and $b_{i}$
($i=0,1,2$) denote two light beams in the different processes. The pump field
between the two FWMs has a $\pi$ phase difference. $\phi$: phase shift to be
measured.}%
\label{fig1}%
\end{figure}

In this paper, we study the phase sensitivity of the SU(1,1) interferometer
for a coherent light together with a squeezed vacuum light as input using the
homodyne detection. For a certain strengths of squeezed and coherent lights,
we describe the optimal parameter condition where the phase sensitivity can
reach the Heisenberg limit. We also compare the homodyne detection with the
intensity detection under this input case, and find that the phase sensitivity
in the homodyne detection is better than that in the intensity detection.
Here, the phase shift studied is not general phase but is sufficiently close
to the optimal phase point \cite{Ou12,Xiang11}.

An SU(1,1) interferometer proposed by Yurke \emph{et al.} \cite{Yurke} is
shown in Fig. \ref{fig1}. In our scheme, the detected variable is amplitude
quadrature $\hat{X}$ other than the photon number $\hat{N}$ which has been
studied in Ref. \cite{Plick,Marino}. Using the amplitude quadrature $\hat{X}$,
the phase sensitivity of the SU(1,1) interferometer is given by%
\begin{equation}
(\triangle\phi)^{2}=\frac{\langle(\triangle\hat{X})^{2}\rangle}{|\partial
\langle\hat{X}\rangle/\partial\phi|^{2}}. \label{eq1}%
\end{equation}

\begin{figure}[ptb]
\centerline{\includegraphics[scale=0.6,angle=0]{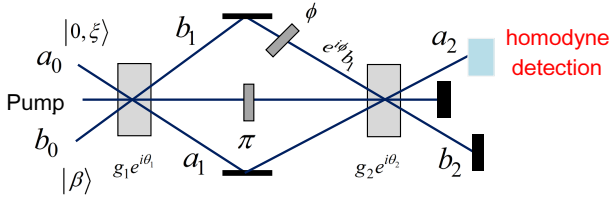}} \caption{(Color
online) A coherent state $|\beta\rangle$ in one input port and a
squeezed-vacuum state $|0,\xi\rangle$ in the other input port of the SU(1,1)
interferometer. The homodyne detection is at the dark port.}%
\label{fig2}%
\end{figure}

We firstly consider a lossless SU(1,1) interferometer with two input ports as
shown in Fig. \ref{fig1}. After the first FWM, one output is retained as a
reference, while the other experience a phase shift process. After the beam
recombine in the second FWM with the reference light, the output fields are
dependent on the phase difference $\phi$ between the two beams. $\hat{a}$
($\hat{a}^{\dagger}$) and $\hat{b}$ ($\hat{b}^{\dagger}$) are the annihilation
(creation) operators for the two mode, respectively. Then the output amplitude
quadrature operator can be written as%
\begin{equation}
\hat{X}\equiv(\hat{a}_{2}+\hat{a}_{2}^{\dagger})/\sqrt{2}. \label{eq2}%
\end{equation}
Through the SU(1,1) transformation \cite{Plick,Marino}, we can obtain%
\begin{align}
\hat{a}_{2}  &  =\mathcal{U}\hat{a}_{0}-\mathcal{V}\hat{b}_{0}^{\dagger
},\label{eq3}\\
\hat{b}_{2}  &  =e^{i\phi}(\mathcal{U}\hat{b}_{0}-\mathcal{V}\hat{a}%
_{0}^{\dagger}), \label{eq4}%
\end{align}
where $\mathcal{U}=\cosh g_{1}\cosh g_{2}+e^{-i\phi}e^{i\left(  \theta
_{2}-\theta_{1}\right)  }\sinh g_{1}\sinh g_{2}$ and $\mathcal{V}%
=e^{i\theta_{1}}\sinh g_{1}\cosh g_{2}+e^{-i\phi}e^{i\theta_{2}}\cosh
g_{1}\sinh g_{2}$, so $\left\vert \mathcal{U}\right\vert ^{2}-\left\vert
\mathcal{V}\right\vert ^{2}=1$. $g_{1}$ ($g_{2}$) and $\theta_{1}$
($\theta_{2}$) describe the strength and phase shift in the process of FWM in
the atomic cell $1$ ($2$), respectively. The balanced situation is $\theta
_{2}-\theta_{1}=\pi$ and $g=g_{1}=g_{2}$ that the second FWM will "undo" what
the first did when the phase shift $\phi$ is $0$.

\begin{figure}[ptb]
\centerline{\includegraphics[scale=0.45,angle=0]{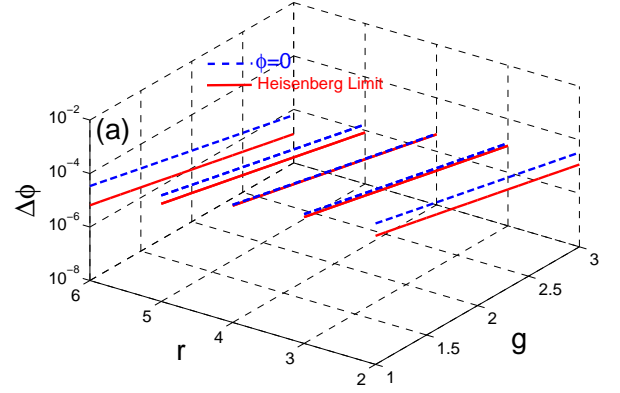}}
\centerline{\includegraphics[scale=0.45,angle=0]{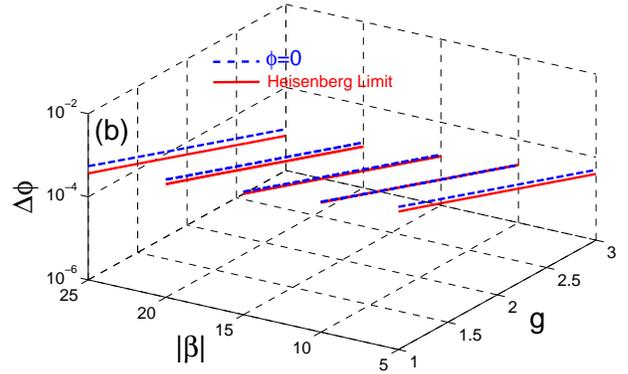}} \caption{(Color
online) (a) The behavior of $\triangle\phi$\ as a function of $g$ for a number
of different values of squeeze strength $r$ with $|\beta|=20$. (b) The
behavior of $\triangle\phi$\ as a function of $g$ for a number of different
values of input coherent state strength $|\beta|$ with $r=3$.}%
\label{fig3}%
\end{figure}

Simply for convenience, next we consider the balanced configuration
($g_{1}=g_{2}=g$ and $\theta_{2}-\theta_{1}=\pi$). Here, we consider a
coherent light $|\beta\rangle$ together with a squeezed vacuum $|0,\xi\rangle$
with $\xi=re^{i\eta}$ as input shown in Fig.~\ref{fig2}, then the noise and
phase sensitivity of the output amplitude quadrature $\hat{X}$ are given by
\begin{align}
&  \langle(\triangle\hat{X})^{2}\rangle=\frac{|\mathcal{V}|^{2}+|\mathcal{U}%
|^{2}[\cosh(2r)-\sinh(2r)\cos(\Theta)]}{2},\label{s1}\\
&  (\triangle\phi)^{2}=\frac{2(\triangle\hat{X}_{\text{s}})^{2}}{\left\vert
\beta\right\vert ^{2}\sinh^{2}(2g)(\cos\Phi)^{2}},\label{s2}%
\end{align}
where $\Phi=\theta_{2}-\theta_{\beta}-\phi-\pi/2$, $\Theta=\eta+2\theta
_{\mathcal{U}}$, $\theta_{\mathcal{U}}$ is given by $\mathcal{U}%
=|\mathcal{U}|\exp(i\theta_{\mathcal{U}})$. When $\Theta=0$, the noise of Eq.
(\ref{s1}) can be $\langle(\triangle\hat{X})^{2}\rangle=$ $e^{-2r}/2$.
Obviously $e^{-2r}<1$ when vacuum squeezed strength $r>0$, therefore the noise
can be reduced to below vacuum noise $1/2$.

At the optimal point $\phi=0$, the best phase sensitivity of our scheme is%
\begin{equation}
(\triangle\phi^{\prime})^{2}=e^{-2r}\frac{1}{\left\vert \beta\right\vert
^{2}\sinh^{2}(2g)},\label{s3}%
\end{equation}
where $\Phi=0$ is used. From Eq.~(\ref{s3}), we find that the sensitivity
$\triangle\phi^{\prime}$ can be improved by the factor $e^{-r}$ from the
squeezed light and by the factor $\sinh(2g)$ from the FWM process. Now, we
compare the phase sensitivity $\triangle\phi^{\prime}$ with the HL of
$\triangle\phi_{\text{HL}}=1/N_{\text{Tot}}$\ where $N_{\text{Tot}}$
($\equiv\langle\hat{a}_{1}^{\dagger}\hat{a}_{1}+\hat{b}_{1}^{\dagger}\hat
{b}_{1}\rangle$) is the total photon number inside the SU(1,1) interferometer,
then
\begin{equation}
\triangle\phi_{\text{HL}}=\frac{1}{\cosh(2g)(\left\vert \beta\right\vert
^{2}+\sinh^{2}r)+2\sinh^{2}g}.\label{s4}%
\end{equation}
When $g=0$, the above result is $\triangle\phi_{\text{HL}}=1/(\left\vert
\beta\right\vert ^{2}+\sinh^{2}r)$ of the traditional MZI\ \cite{Caves81}. The
performances of phase sensitivity in our scheme are shown in Fig.
\ref{fig3}(a) and (b) at the optimal point $\phi=0$. The behavior of
$\triangle\phi$ as a function of $g$ for a number of different values of
squeeze strength $r$ and of input coherent state intensity $\left\vert
\beta\right\vert $ is shown in Fig. \ref{fig3}(a) and Fig. \ref{fig3}(b),
respectively. From Fig. \ref{fig3}(a), we obtain that for a certain intensity
of coherent light input, the phase sensitivity is not always better with
increasing squeezed strength $r$. In a similar way, for a certain intensity of
squeezed light input, the phase sensitivity is not always better with
increasing input coherent light strength, which is shown in Fig. \ref{fig3}(b).

With Eqs. (\ref{s3}) and (\ref{s4}), the optimal condition to reach HL is%
\begin{equation}
\left\vert \beta\right\vert \simeq\frac{e^{r}}{2}\tanh(2g).\label{s5}%
\end{equation}
Therefore, when the condition $\left\vert \beta\right\vert \simeq e^{r}%
\tanh(2g)/2$ is met between the input coherent state, the input squeezed
vacuum state and the FWM process, the phase sensitivity can reach HL.

\begin{figure}[ptb]
\centerline{\includegraphics[scale=0.55,angle=0]{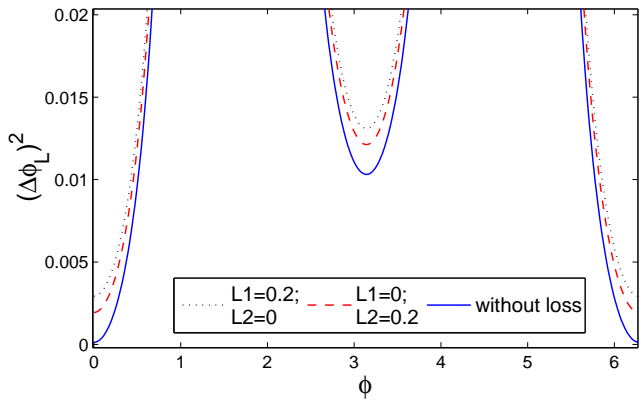}} \caption{(Color
online) The phase sensitivity as a function of $\phi$ with loss. The dashed
line indicates the case of the outside loss $L_{2}=0.2$ without the internal
loss $L_{1}=0$, and the dotted line shows the case of the internal loss
$L_{1}=0.2$ without the outside loss $L_{2}=0$ while the solid line indicates
the ideal lossless case $L_{1}=L_{2}=0$. The parameters are as follows:
$|\beta|=10$, $\theta_{\beta}$=$\pi/2$, $r=2$, $g=0.5$.}%
\label{fig4}%
\end{figure}

As has been previously pointed out that the loss is the limiting factor in
precision measurement \cite{Ou12,loss1,loss2}. Next, we investigate the
effects of photon losses on the phase sensitivity of this coherent $\otimes$
squeezed-vacuum light as input case. Firstly we consider that both arms of the
interferometer have the same internal losses $L_{1}$ and outside loss $L_{2}$.
After taking account into these losses we introduce that $\langle\hat
{X}_{\text{L}}\rangle$, compared with Eq. (\ref{eq2}), can be written as%
\begin{equation}
\langle\hat{X}_{\text{L}}\rangle=\langle\hat{X}\rangle\sqrt{1-L_{1}}%
\sqrt{1-L_{2}},\label{301}%
\end{equation}
where the subindex L means that the losses are considered. As shown in Fig.
\ref{fig2}, in the balanced case, considering the losses the sensitivity is
given by%
\begin{align}
(\triangle\phi_{\text{L}})^{2} &  =(\triangle\phi)^{2}+\frac{1}{\left\vert
\beta\right\vert ^{2}\sinh^{2}(2g)}\nonumber\\
&  \times\frac{\cosh(2g)L_{1}(1-L_{2})+L_{2}}{(1-L_{1})(1-L_{2})\cos^{2}\Phi
},\label{305}%
\end{align}
where $\triangle\phi$ is from Eq. (\ref{s2}), and the second term on the
right-hand side is the extra noise term from loss. 

The inside losses make more
impact on the sensitivity of phase measurement than the outside losses, which
can be seen from the term $\cosh(2g)L_{1}(1-L_{2})+L_{2}]/(1-L_{1})(1-L_{2})$
due to $\cosh(2g)$. When $L_{1}\neq0$ and $L_{2}=0$, the effect is slight
larger than that of the case $L_{1}=0$ and $L_{2}\neq0$ due to $\cosh(2g)>1$.
We calculate the internal loss by making $L_{2}=0$ in Eq. (\ref{305}), then it
can be written as%
\begin{equation}
(\triangle\phi_{\text{L}})^{2}=(\triangle\phi)^{2}+\frac{L_{1}\cosh
(2g)}{(1-L_{1})\left\vert \beta\right\vert ^{2}\sinh^{2}(2g)\cos^{2}\Phi
}.\label{306}%
\end{equation}
Similarly, when only consider the outside loss that letting $L_{1}=0$, such
that,
\begin{equation}
(\triangle\phi_{\text{L}})^{2}=(\triangle\phi)^{2}+\frac{L_{2}}{(1-L_{2}%
)\left\vert \beta\right\vert ^{2}\sinh^{2}(2g)\cos^{2}\Phi}.\label{307}%
\end{equation}
The comparison of Eq. (\ref{306}) and Eq. (\ref{307}) is shown in Fig.
\ref{fig4}. Obviously, one can see that the effect of the internal loss is
slightly greater than the outside loss.

\begin{figure}[ptb]
\centerline{\includegraphics[scale=0.47,angle=0]{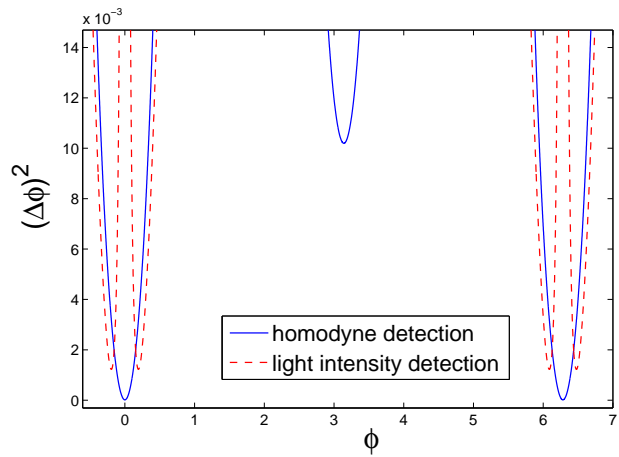}}\caption{(Color
online) The phase sensitivity comparison between light intensity detection and
the homodyne detection under the same condition for a coherent light together
with a squeezed vacuum state input. The parameters are as follows: $g=1$,
$r=2$, $|\beta|=10$, $\eta=0$.}%
\label{fig5}%
\end{figure}

Finally, we will give a brief comparison between amplitude quadrature
detection $\hat{X}$ and output light intensity detection $\hat{N}$
($\equiv\hat{n}_{a_{2}}+\hat{n}_{b_{2}}$) which has been discussed in Ref.
\cite{Plick,Marino}. Using the light intensity\ detection, the sensitivity of
SU(1,1) interferometer is \cite{Plick,Marino}%
\begin{equation}
(\triangle\phi^{\text{N}})^{2}=\frac{\langle(\triangle\hat{N})^{2}\rangle
}{|\partial\langle\hat{N}\rangle/\partial\phi|^{2}},\label{401}%
\end{equation}
where the superscript N means light intensity detection. We consider the case
that a coherent light $|\beta\rangle$ together with a squeezed vacuum
$|0,\xi\rangle$ input, without losses, the phase sensitivity of intensity
detection is given by
\begin{align}
(\Delta\phi_{\text{s}}^{\text{N}})^{2} &  =\frac{(|\beta|^{2}+1)^{2}}%
{(|\beta|^{2}+1+\sinh^{2}r)^{2}}(\Delta\phi_{\text{c}}^{\text{N}}%
)^{2}+A,\label{403}\\
(\triangle\phi_{\text{c}}^{\text{N}})^{2} &  =\frac{|\beta|^{2}(|\mathcal{U}%
|^{2}+|\mathcal{V}|^{2})^{2}+4|\mathcal{U}|^{2}|\mathcal{V}|^{2}(|\beta
|^{2}+1)}{16(|\beta|^{2}+1)^{2}\sin^{2}(\phi)\sinh^{4}(g)\cosh^{4}(g)},\\
A &  =\dfrac{\Lambda}{16(|\beta|^{2}+1+\sinh^{2}r)^{2}\sin^{2}(\phi)\sinh
^{2}g\cosh^{2}g},\\
\Lambda &  =(|\mathcal{U}|^{2}+|\mathcal{V}|^{2})^{2}(1+\cosh^{2}r)\sinh
^{2}r\nonumber\\
&  +4|\mathcal{U}|^{2}|\mathcal{V}|^{2}(1+2|\beta|^{2})\sinh^{2}r\nonumber\\
&  +[4\mathcal{U}^{2}(\mathcal{V}^{\ast})^{2}\beta^{2}\cosh r\sinh re^{i\eta
}+c.c.],
\end{align}
where $\Delta\phi_{\text{s}}^{\text{N}}$ and $\triangle\phi_{\text{c}%
}^{\text{N}}$\ are the phase sensitivities of coherent $\otimes$ vacuum light
as input\ and coherent $\otimes$ squeezed-vacuum light as input, respectively.
Under the same condition, the phase sensitivities by the homodyne detection
and intensity detection are shown in Fig. \ref{fig5}. Obviously, one can
obtain that the performance of the homodyne detection is slightly better than
that of the intensity detection.

In summary, we investigated the phase sensitivity of the SU(1,1)
interferometer for a coherent state and a squeezed vacuum state as input. We
obtain that the the relative phase sensitivity is not always better with
increasing the input squeezed strength or increasing the input coherent light
intensity, and give out the optimal condition to reach the HL. For the loss
effect, the inside losses make more impact on the sensitivity of phase
measurement than the outside losses. Compared to the intensity detection, our
scheme shows a slightly better phase sensitivity in precision phase measurement involving
squeezing state input case.

We thank Prof. Jietai Jing for his helpful discussions. This work was
supported by the National Basic Research Program of China (973 Program) under
Grant No.~2011CB921604 and No.~11234003 (W.Z.), the National Natural Science
Foundation of China under Grant No.~11129402 (Z.Y.O.), the National Natural
Science Foundation of China under Grant No.~11004059 (C.H.Y.). \newline Email:
$^{\ast}$chyuan@phy.ecnu.edu.cn;\newline$~~~~~~~~~~~^{\dag}$wpzhang@phy.ecnu.edu.cn.

\end{document}